\documentclass[preprint2]{emulateapj}
\usepackage{subfigure}

\slugcomment{accepted to ApJL}

\shorttitle{BRAVA-RR}
\shortauthors{Kunder et al.}

\begin{document}

\title{A high-velocity bulge RR Lyrae variable on a halo-like orbit}

\author{Andrea Kunder\altaffilmark{1},
R.~M.~Rich\altaffilmark{2},
K.~Hawkins\altaffilmark{3},
R.~Poleski\altaffilmark{4,5}, 
J.~Storm\altaffilmark{1},
C.~I.~Johnson\altaffilmark{6},
J.~Shen\altaffilmark{7},
Z.-Y.~Li\altaffilmark{7},
M.~J.~Cordero\altaffilmark{8},
D.~M.~Nataf\altaffilmark{9},
G.~Bono\altaffilmark{10,11},
A.~R.~Walker\altaffilmark{12},
A.~Koch\altaffilmark{13},
R.~De Propris\altaffilmark{14},
A. Udalski\altaffilmark{5},
M.~K.~Szyma{\'n}ski\altaffilmark{5},
I. Soszy{\'n}ski\altaffilmark{5},
G. Pietrzy{\'n}ski\altaffilmark{5,15},
K. Ulaczyk\altaffilmark{5,16},
{\L}. Wyrzykowski\altaffilmark{5},
P. Pietrukowicz\altaffilmark{5},
J. Skowron\altaffilmark{5},
S. Koz{\l}owski\altaffilmark{5},
P. Mr{\'o}z\altaffilmark{5}
}
\altaffiltext{1}{Leibniz-Institut f\"{u}Ÿr Astrophysik Potsdam (AIP), An der Sternwarte 16, D-14482 Potsdam, Germany}
\altaffiltext{2}{Department of Physics and Astronomy, University of California at Los Angeles, Los Angeles, CA 90095-1562, USA}
\altaffiltext{3}{Institute of Astronomy, Madingley Road, Cambridge CB3 0HA, UK}
\altaffiltext{4}{Department of Astronomy, Ohio State University, 140 W. 18th Ave., Columbus, OH 43210, USA}
\altaffiltext{5}{Warsaw University Observatory, Al. Ujazdowskie 4, 00-478 Warszawa, Poland}
\altaffiltext{6}{Harvard-Smithsonian Center for Astrophysics, Cambridge, MA 02138}
\altaffiltext{7}{Key Laboratory for Research in Galaxies and Cosmology, Shanghai Astronomical
Observatory, Chinese Academy of Sciences, 80 Nandan Road, Shanghai 200030, China}
\altaffiltext{8}{Astronomisches Rechen-Institut: Zentrum f\"{u}r Astronomie, Mönchhofstr. 12-14, 69120 Heidelberg Germany}
\altaffiltext{9}{Research School of Astronomy and Astrophysics, The Australian National University, Canberra, ACT 2611, Australia}
\altaffiltext{10}{Dipartimento di Fisica, Universita di Roma Tor Vergata, vi a Della Ricerca Scientifica 1, 00133, Roma, Italy}
\altaffiltext{11}{INAF,  Rome  Astronomical  Observatory,  via  Frascati  33, 00040, Monte Porzio Catone, Italy}
\altaffiltext{12}{Cerro Tololo Inter-American Observatory,  National Optical Astronomy Observatory, Casilla 603, La Serena, Chile}
\altaffiltext{13}{Landessternwarte, Zentrum f\"{u}r Astronomie der Universit\"{a}t Heidelberg, K\"{o}nigstuhl 12, D-69117 Heidelberg, Germany}
\altaffiltext{14}{Finnish Centre for Astronomy with ESO (FINCA), University of Turku, Finland}
\altaffiltext{15}{Universidad de Concepci{\'o}n, Departamento de Astronomia, Casilla 160-C, Concepci{\'o}n, Chile}
\altaffiltext{16}{Department of Physics, University of Warwick, Gibbet Hill Road, Coventry, CV4 7AL, UK}

\begin{abstract}
We report on the RR Lyrae variable star, MACHO~176.18833.411, located toward the
Galactic bulge and observed within
the data from the ongoing Bulge RR Lyrae Radial Velocity Assay (BRAVA-RR), which
has the unusual radial velocity of $-$372~$\pm$~8~km~s$^{-1}$ and true space velocity
of $-$482~$\pm$~22~km~s$^{-1}$ relative to the Galactic rest frame. Located less than 1~kpc from 
the Galactic center and toward a field at ($l$,$b$)=(3,$-$2.5),
this pulsating star has properties suggesting it belongs
to the bulge RR Lyrae star population yet a velocity indicating it is abnormal, at least
with respect to bulge giants and red clump stars.  We show that this star is most
likely a halo interloper and therefore suggest that halo contamination is not insignificant
when studying metal-poor stars found within the bulge area,
even for stars within 1~kpc of the Galactic center.  We discuss the possibility that 
MACHO 176.18833.411 is on the extreme edge of the bulge RR Lyrae radial velocity distribution, 
and also consider a more exotic scenario in which it is a runaway star moving through the Galaxy.
\end{abstract}


\section{Introduction}
It is well known that within the Galactic bulge at longitudes $\rm|$$l$$\rm|$$<$10$^\circ$,
there is a bar-like structure with 
a bar angle in the range 20 -- 30$^\circ$, traced by numerous old stellar population 
probes (e.g., red clump giants, Stanek 1994, Miras, Groenewegen \& Blommaert 2005).  
This bar/bulge is a rotating Box/Peanut (B/P) structure with a X-shape
protrusion, made up largely of old and metal-rich
stars ($\sim$10 Gyr, $\rm [Fe/H]$ falling between $-$0.5 and +0.5~dex, 
e.g., Johnson et al. 2013; Gonzalez et~al. 2012; Wegg \& Gerhard 2013).

There is speculation that the Milky Way also has an older, more spheroidal 
bulge population, and perhaps the greatest possibility of uncovering
such a component would be within the most metal-poor bulge stars.
One example of a metal-poor bulge population is the RR Lyrae stars (RRLs),
and over the past few years, $\sim$38~000 RRLs toward the bulge have been identified
from photometric surveys \citep[e.g.,][]{soszynski14}. 
These RRLs are thought to exhibit a small metallicity spread and are centered 
around $\rm [Fe/H]$ = $-$1~dex \citep{walker91, kunderfe08, soszynski14}.  The
small $\rm [Fe/H]$ spread may suggest that RRLs trace a more ancient 
stellar population than the majority of bulge RGB and red clump stars, which are 
more metal-rich, on average, and likely do not evolve to become RRLs \citep[e.g.][]{lee92, walker91}.  

Recent photometric studies have reached different conclusions regarding the
relationship of the bulge RRLs and the bar/bulge. 
\citet{dekany13} combined optical and infrared photometry of $\sim$8000 ss
OGLE-III discovered RRLs to find that unlike the red clump giants (RCGs), RRLs do not trace a strong bar.  
Instead, they have a more spheroidal, centrally concentrated
distribution, indicating that the RRLs belong to a classical bulge that has 
co-evolved with the bar \citep[e.g.,][]{saha13}.  

In contrast, using 28~000 RRLs from the more spatially extended
OGLE-IV bulge sample, \citet{pietrukowicz14} assert
that the RRLs trace closely the barred structure formed of RCGs, and hence
that the bulge RRLs are in the same gravitational potential together with the more massive Galactic bar.

Missing still are the kinematics of the bulge RRLs, which can resolve this discrepancy and 
provide an understanding of the origin of the old, metal-poor bulge component.  
The last published paper on bulge RRLs radial velocities, \citet{gratton87}, used a sample
of 17 RRLs to conclude that the kinematic properties of RRLs in Baade's Window 
are similar to that of the Miras, M-giants, K-giants, OH/IR sources and planetary 
nebulae of the Galactic bulge.  

In this paper, we report on a high-velocity RRL found serendipitously in the 
Bulge RR Lyrae Radial Velocity Assay (BRAVA-RR) to have $v_{r}$=$-$372 km s$^{-1}$, 
which is well above the typical speed of the stars one might expect to find in the bulge.
High-velocity stars are intriguing in part because they can provide insight to the mechanisms that produce 
their velocities.  The origin of high-velocity stars can also provide useful information about the environments 
from which they are produced.  Here we investigate 
the cause for the high-velocity of MACHO~176.18833.411 to discern whether it is consistent with stars in the 
Galactic bulge and what this suggests about the formation of the Galaxy.

\section{Observations and Radial Velocity}
MACHO~176.18833.411 was originally catalogued by the MACHO survey as a
fundamental mode RRL \citep{kunder08}\footnote{This RRL is also designated
OGLE-BLG-RRLYR-10353.}.
The star was one of the $\sim$100 RRLs surveyed spectroscopically 
as part of BRAVA-RR (NOAO PropID: 2014A-0143; PI: A. Kunder) in a field at ($l$,$b$)=(3,$-$2.5) 
using the AAOmega multifiber spectrograph 
on the Anglo-Australian Telescope (AAT).  We observed this star
twice on June 21, 2014, separated in time by 7 hours, and
the observations were taken in dual beam mode centered on 
8600\AA, with the 580V and 1700D gratings to probe the Calcium Triplet.  This covers the 
optical window from about 8300\AA~to 8800\AA~at a resolution of R$\sim$10,000. 

The data were reduced using the automated pipeline supplied by AAOmega, 2DFDR,
and the spectra were cross-correlated using the IRAF cross-correlation routine, {\tt xcsoa}.
Four Bulge RAdial Velocity Assay (BRAVA) stars \citep{kunder12} observed
with the same setup were selected as radial velocity standard stars.  
Due to less than optimal weather conditions, the signal-to-noise ratio is $\sim$10 (see 
spectrum in Figure~\ref{lc}), and the consistency of our
velocity result is 8~km~s$^{-1}$, in agreement with the errors reported by {\tt xcsao} for each 
individual measurement.

Figure~\ref{lc} (middle) shows the pulsation curve using the radial velocity template and scaled by its $V$-amplitude
as outlined from \citet{liu91}.  The two radial velocity measurements were folded by the known period 
to find the radial velocity as a function of phase.  Both measurements fit the radial velocity template 
well for a RRL with a line-of-sight radial velocity, $V_{los}$ of $-$372~km~s$^{-1}$.  
It is noteworthy that we did not adjust the template in phase, which indicates that the OGLE
time of maximum brightness is reliable for this star.  

\section{The Properties of MACHO~176.18833.411}
\begin{figure*}
\centering
\subfigure{
\includegraphics[height=5cm]{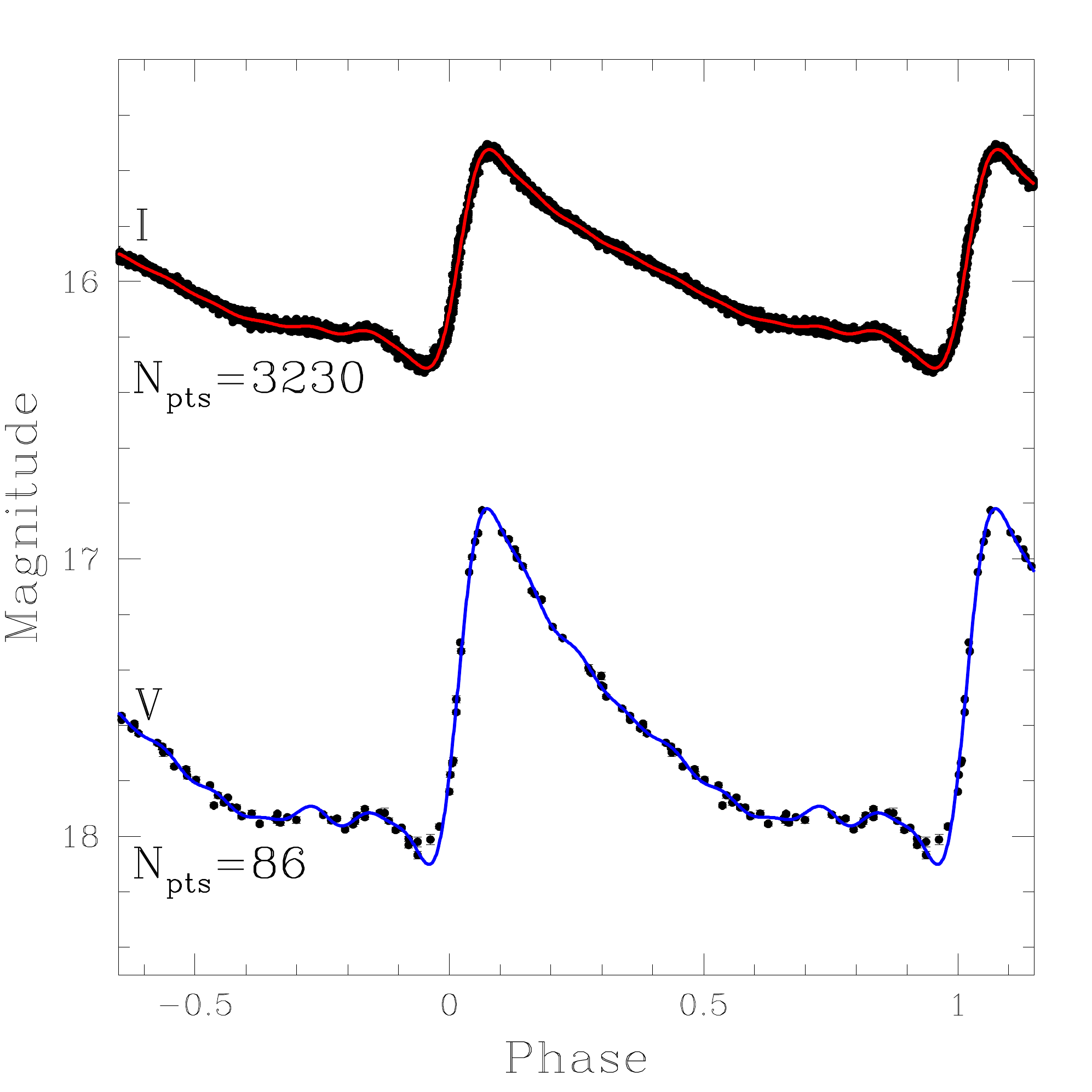}
}
\quad
\subfigure{
\includegraphics[height=5cm]{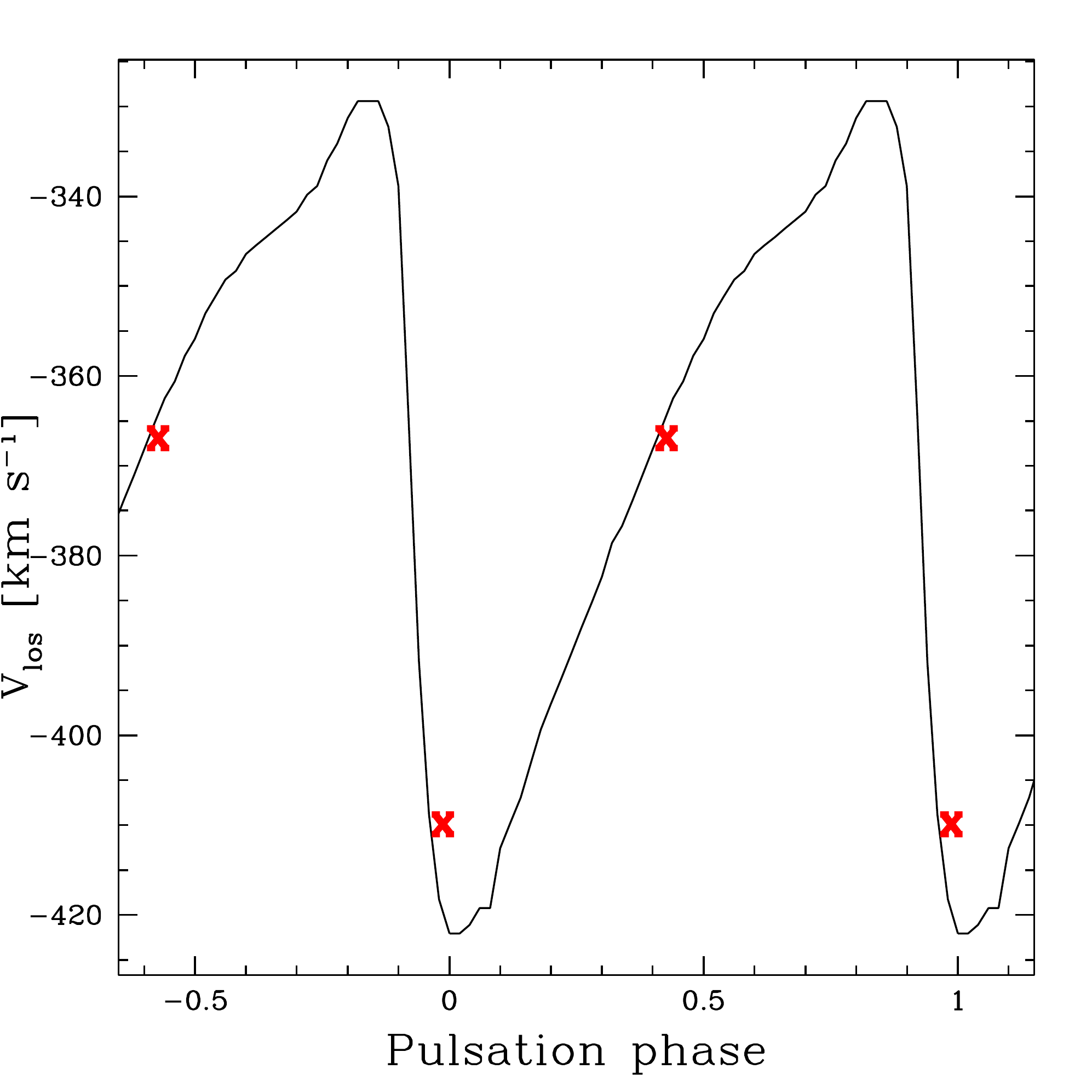} 
}
\subfigure{
\includegraphics[height=5cm]{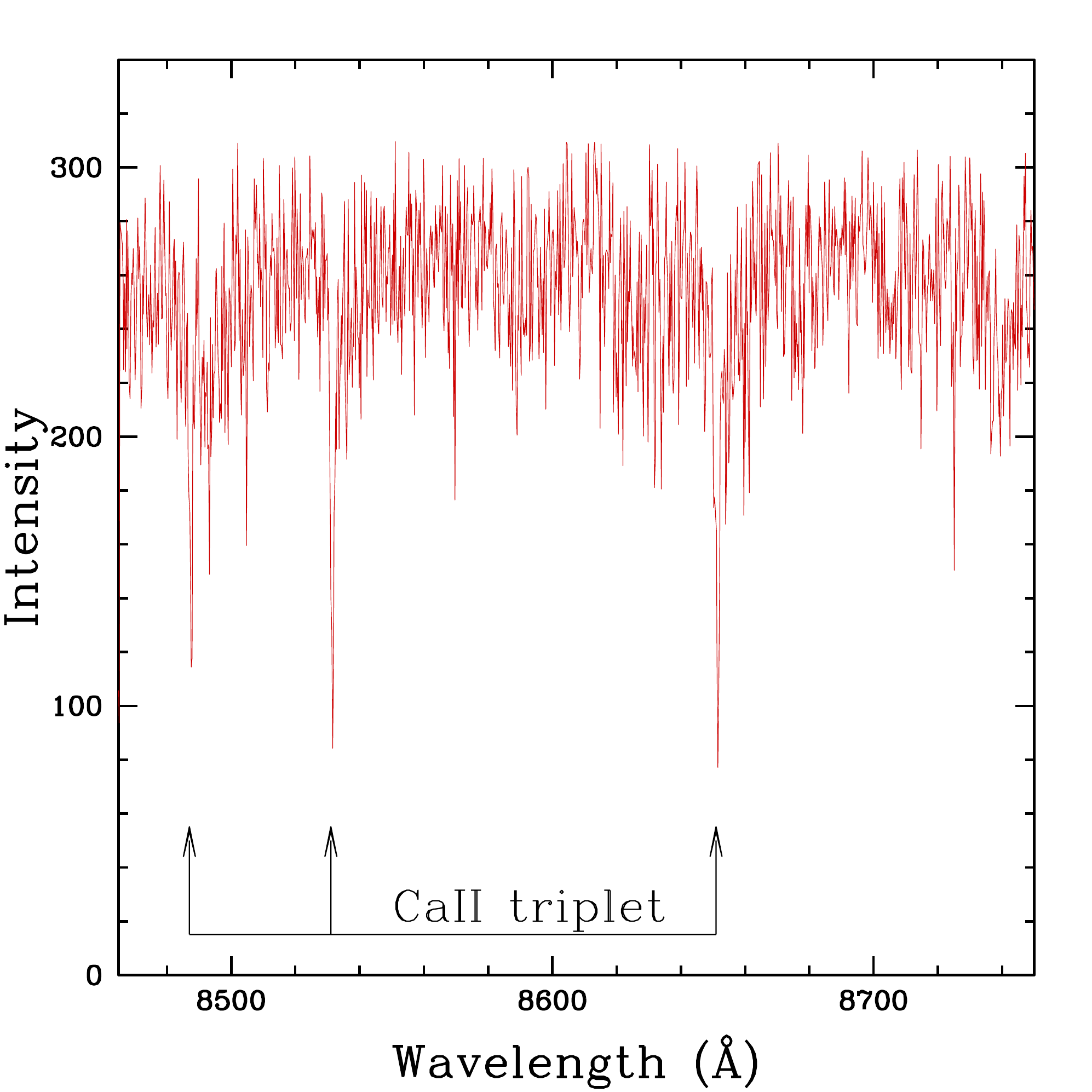} 
}
\caption{{\sl Left:} Phased $VI$ OGLE-IV light curves of MACHO~176.18833.411 with the Fourier fit over-plotted.
{\sl Middle:} Line-of-sight radial velocity versus pulsational phase for our two observations, over-plotted on a
fundamental mode RRL radial velocity template and scaled by its $V$-amplitude \citep{liu91}.
{\sl Right:} Wavelength-calibrated spectrum of MACHO~176.18833.411. The CaT lines are labeled 
for reference.
}
\label{lc}
\end{figure*}

The OGLE-IV catalogue of RRLs provides $V$- and $I$-band light curves of their stars, with
the photometric observations 
spanning from March 2010 to October 2013 \citep{soszynski14}. 
The optical light curve for MACHO~176.18833.411 from the OGLE-IV observations 
is shown in Figure~\ref{lc} (left), and its most important properties are summarised in Table~\ref{properties}.

To find the distance to the RRL, we first use the mean-flux magnitude as listed by OGLE-IV. 
Next, the E$\hbox{\it (V--I)\/}$ color excess along the stars line of sight is calculated from the observed 
\hbox{\it V--I\/} color at minimum light, $\hbox{\it (V--I)\/}_{min,obs}$  \citep[see e.g., ][]{guldenschuh05},
by carrying out a Fourier fit to the $V$ and $I$-band light-curves.  

The extinction, $A_I$ can then be derived using:
\begin{equation}
A_I = 0.7465 E(V-I) + 1.37 E(J-K),
\end{equation}
as introduced in \citet{nataf13}.  Here, E$\hbox{\it (J--K)\/}$ was taken from the bulge reddening 
maps of \citet{gonzalez12}.  
Finally, it has been shown that RRLs $V$-band light curves and the phase difference of their Fourier 
decomposition can be used to estimate photometric metallicities to an accuracy of 0.2~dex\citep{jurcsik96},
although with high-quality light curves, the fitting accuracy can be $\sim$0.12 dex \citep{kovacs05}.  
This relation is well-calibrated over the metallicity range from $\sim$$-$2.0~dex to $\sim$0~dex.
Applying this method to the OGLE $V$-band light curve, the metallicity ($\rm [Fe/H]$) 
of MACHO~176.18833.411 is estimated and placed on the \citet{carretta09} 
metallicity scale.
Using the recalibration of the RRL luminosity scale by \citet{catelancortes08}, 
MACHO~176.18833.411 has an absolute magnitude of $M_V$=0.58$\pm$0.13 mag.
Similarly, using a quadratic relation between RRLs absolute magnitude
and metallicity from Bono, Caputo, \& di Criscienzo (2007), we find
 $M_V$=0.61$\pm$0.08~mag, 
where 0.08 is a reasonable error in the RRLs absolute magnitude-$\rm[Fe/H]$ zero-point
calibration.  
The \citet{benedict11} $M_V$--$\rm [Fe/H]$ relation, however, 
results in $M_V$=0.43$\pm$0.07, indicating that the level of agreement between 
independent $M_V$ measurements is as large as $\sim$0.2 mag.  
A systematic uncertainty of 0.2~mag $M_V$ is factored into the uncertainty in 
the estimated distance, and does not change the results significantly.
The apparent magnitude, reddening and absolute magnitude of MACHO~176.18833.411 
lead to a distance of $(m-M)_0$=14.29$\pm$0.13~mag.

In the OGLE-III catalog, \citet{soszynski11} marked $\sim$400 stars as possessing proper motions 
relative to mean motion of the bulge that are large enough to be easily detected.
MACHO~176.18833.411 is included in this list.  To determine its proper motion,
the centroids of stars were measured on each OGLE-IV \citep{udalski15} image separately. All measured 
centroids were transformed to the common grid using positions of bright red giants $i.e.,$ stars belonging to 
the bulge, not the disk population. The transformed positions were fitted with a model that takes into account 
proper motion as well as differential refraction effects \citep[see][for details]{poleski13}.

With our proper motion, the space velocity and position vector can be resolved, and the 6D position and velocity 
information for MACHO~176.18833.411 is given in Table~1.  For this convention, the Sun's orbital velocity vector
$v_{\sun}$= [$U_{\sun}$,$V_{\sun}$,$W_{\sun}$]=[14.0,12.24,7.25] km~s$^{-1}$, $V_{LSR}$=220~km~s$^{-1}$
and position = [8.28,0,0] kpc.  The derived true space velocity is 
$-$482~$\pm$~22~km~s$^{-1}$ relative to the Galactic rest frame.

\begin{table}
\caption{Properties of MACHO~176.18833.411}
\label{properties}
\begin{tabular}{ll} \hline
\hline
$\alpha$ (J2000) & 18:00:13.08  \\
$\delta$ (J2000) & $-$27:15:39.1  \\ 
$l$ & 3.0795  \\  
$b$ & $-$1.9209 \\  
$\hbox{\it $<$V$>$\/} $ & 17.586~mag  \\  
$\hbox{\it $<$I$>$\/} $ & 15.969~mag   \\  
$V$-amp & 1.27 $\pm$0.05~mag \\  
Period  & 0.51521996 $\pm$ 0.00000005 d  \\ 
Time of max brightness & 2 456 000.21386 \\
$\rm [Fe/H]$  &  $-$1.62$\pm$0.2 (from light curve) \\ 
($\hbox{\it V--I\/})_{min}$ &  1.79$\pm$0.01 \\ 
E($\hbox{\it V--I\/})_{min}$  &  1.21$\pm$0.03 \\ 
$A_I$  &  1.49 \\
$(m-M)_0$ & 14.29$\pm$0.13 mag  \\ 
Heliocentric Distance & 7300$\pm$600 pc  \\ 
Galactocentric Distance  & 850 pc  \\ 
$V_{los} $ & $-$372 $\pm$ 8~km~s$^{-1}$\\
$V_{GRV} $ & $-$350 $\pm$ 8~km~s$^{-1}$ \\
$\mu_{\alpha*cos(\delta)} $ & 8.08 $\pm$ 0.20 $\pm$ 0.40 mas~yr$^{-1}$ \\
$\mu_{\delta} $ & 4.41 $\pm$ 0.20 $\pm$ 0.40 mas~yr$^{-1}$ \\ 
X &  $-$1.04 $\pm$ 0.56 kpc \\  
Y &  0.39 $\pm$ 0.03 kpc \\ 
Z &  $-$0.243 $\pm$ 0.02 kpc \\ 
U &  $-$377 $\pm$ 20~km~s$^{-1}$ \\
V &  $-$262 $\pm$ 26~km~s$^{-1}$ \\
W &  $-$147 $\pm$ 20~km~s$^{-1}$ \\
\end{tabular}
  \end{table}

%
%
\section{Discussion}
\subsection{Possible explanations for the high-velocity star}
BRAVA-RR currently has surveyed only 94 RRLs, including MACHO~176.18833.411.
Is such a bulge star therefore really anomalous?
To give us some idea if this RRL in fact belongs to the bulge, we integrated its orbit through 
an assumed Galactic potential which is a sum of the 
potential of a logarithmic halo, Miyamoto-Nagai disk, and a Hernquist bulge, as in \citet{hawkins15}.
Uncertainties in the orbital integrations were estimated by a Monte Carlo approach, where 
the initial conditions were varied to within their uncertainties over 100 orbital 
integrations \citep[see][for details]{hawkins15}.  

\begin{figure*}
\centering
\subfigure{
\includegraphics[height=6.4cm]{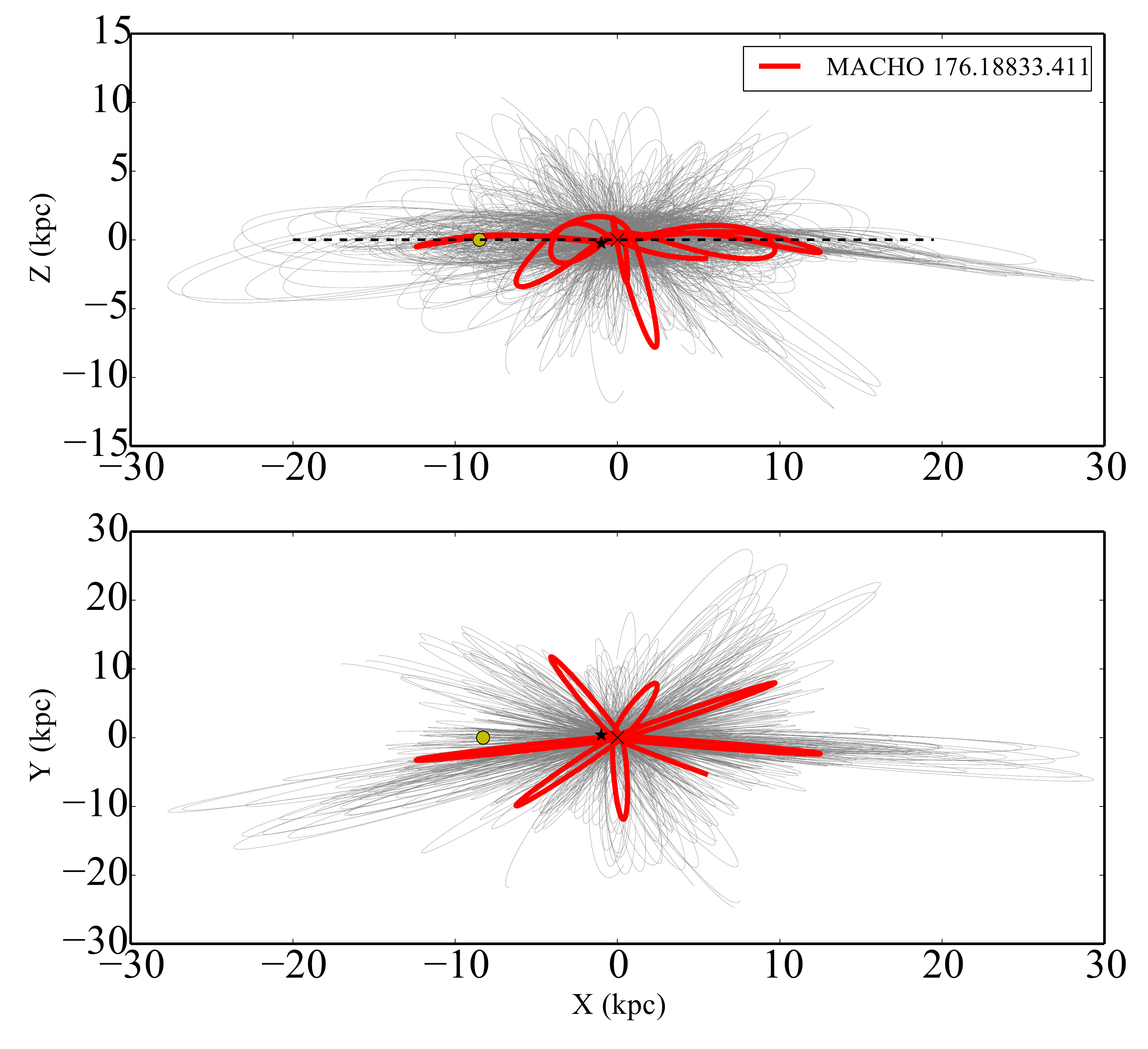}
}
\quad
\subfigure{
\includegraphics[height=6.4cm]{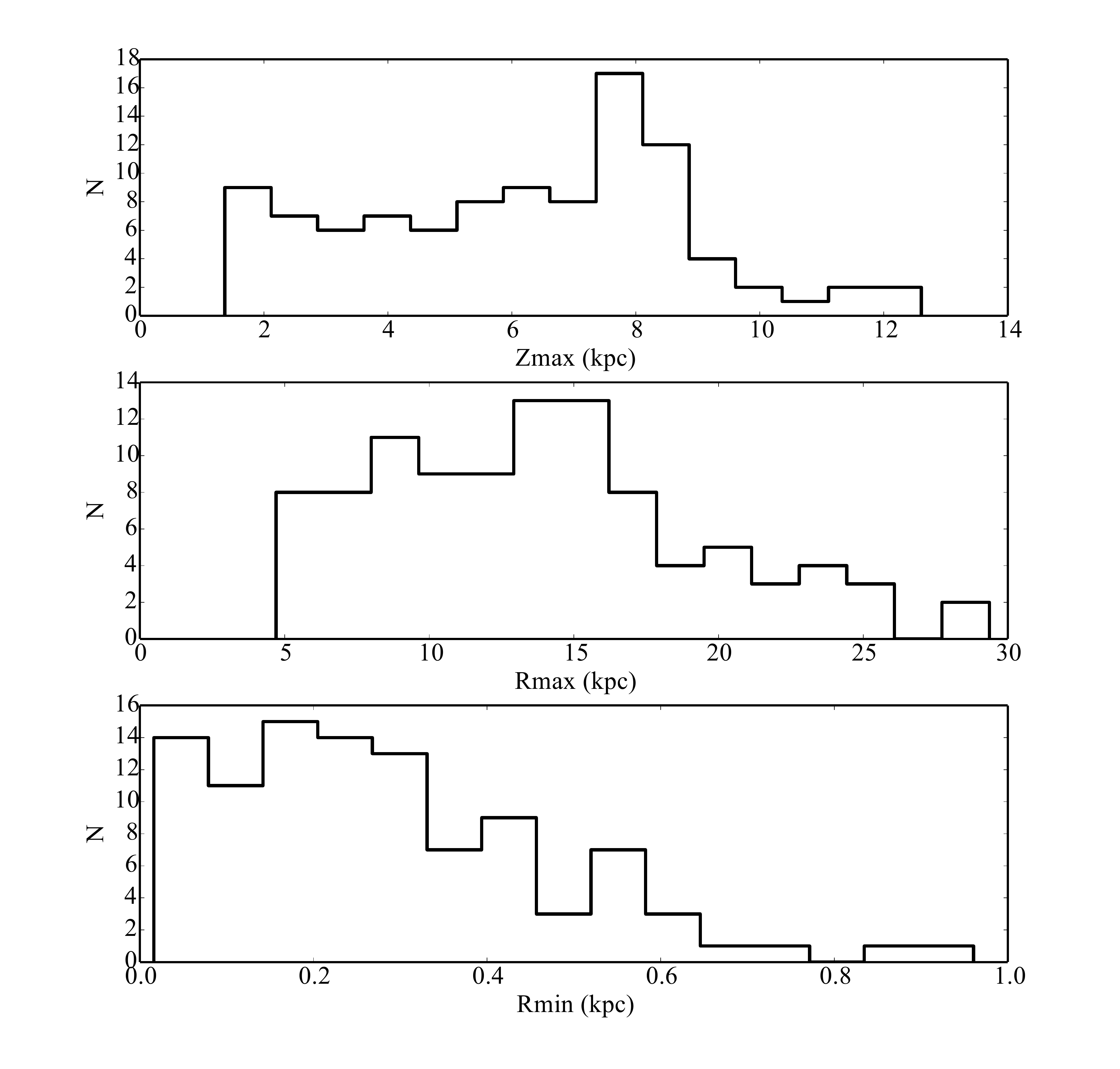}
}
\caption{ {\sl Left:} A 1 Gyr orbital integration for MACHO~176.18833.411 (thick line).
The open circle represents the Sun, the black 'x' represents the Galactic center and the black
asterisk represents the current position of the RRL.  The thin grey
lines are 100 draws of the orbital integration to illustrate
the uncertainty of the orbit.  MACHO~176.18833.411 has an orbit consistent with
that of the halo.
{\sl Right:} The distribution of $R_{min}$, $R_{max}$ and $Z_{max}$ of the 100 orbital 
draws for our RRL. 
}
\label{orbit}
\end{figure*}

The orbital integration over the past 1~Gyr is shown in Figure~\ref{orbit}, as is 
the distribution of the maximum distance from the Galactic plane, $\rm Z_{max}$, and the 
minimum and maximum distance from the Galactic center for 100 orbital draws.
Although MACHO~176.18833.411 is 
currently close to or in the bulge, its orbit clearly suggests it is not confined to the bulge.  
The escape velocity at the radius of the bulge is $\sim$650 km~s$^{-1}$;
this means that a star with the high velocity of MACHO~176.18833.411 
can still be inside the bulge.  However, it is clear that MACHO~176.18833.411 
spends most of its time well outside the radius of the formal bulge/bar structure. 

\begin{figure}[htb]
\includegraphics[width=1\hsize]{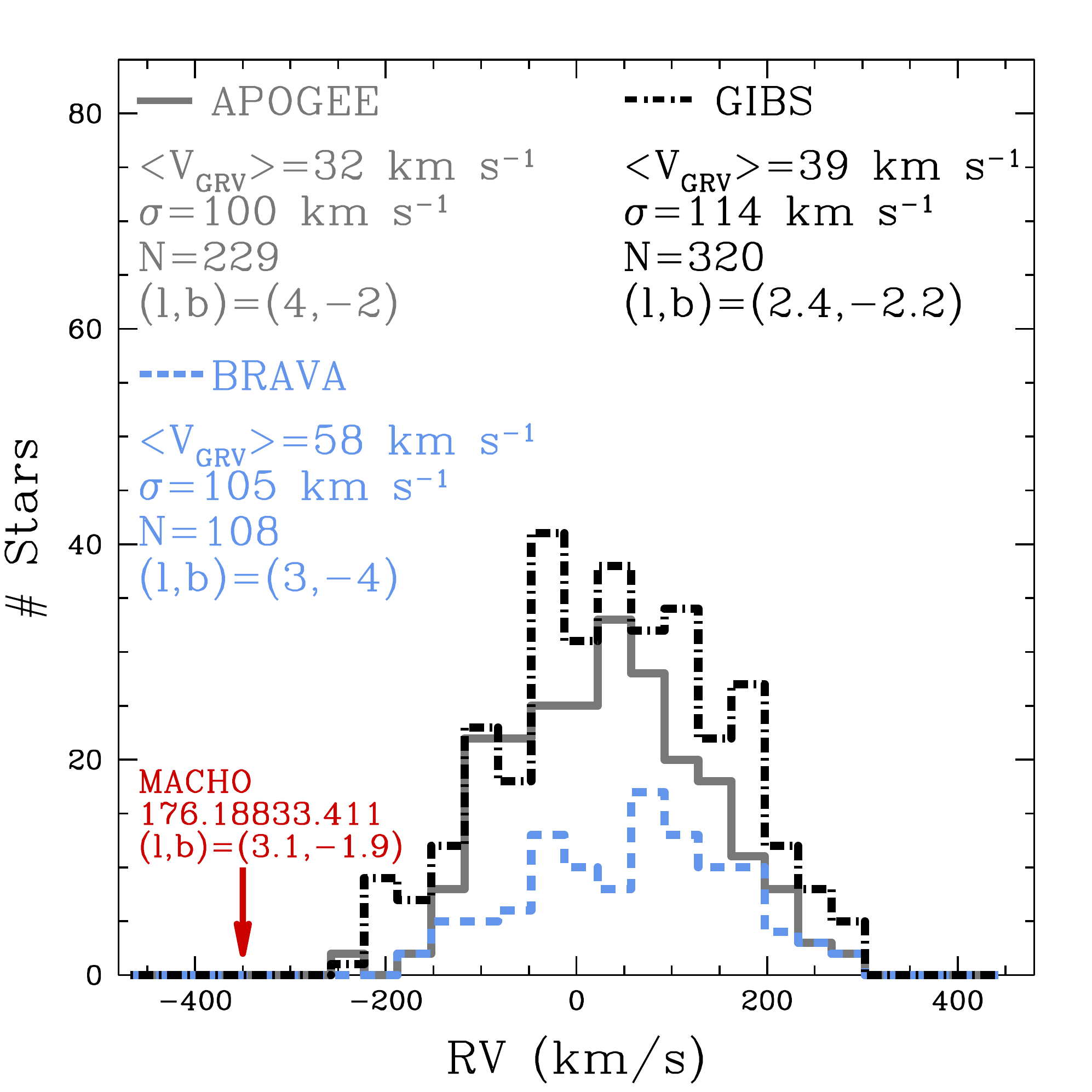}
\caption{The Galactocentric velocity ($V_{GRV}$) distribution of APOGEE giants (solid), 
BRAVA giants (dashed) and GIBS RCGs (dashed-dot) situated spatially close to
MACHO~176.18833.411 (arrow).
The mean GRV, dispersion of the $V_{GRV}$ and total number of stars are
also given.  }
\label{apogee}
\end{figure}

We next address how likely it is for a bulge star to have such a negative radial velocity.  
The Galactocentic radial velocity, $V_{GRV}$, distribution of 229 giants toward 
the bulge at ($l$,$b$)=(4,$-$2) and 320 RCGs toward the bulge at ($l$,$b$)=(2.4,$-$2.2) 
is shown in Figure~\ref{apogee}.  The velocities of the giants are taken from 
APOGEE (the Apache Point Observatory Galactic Evolution Experiment), 
which is part of Sloan Digital Sky Survey III, and we
analyse data from data release 12 (DR12, Alam et~al. 2015).  The velocities of
the RCGs are taken from the GIRAFFE Inner Bulge Survey (GIBS, Zoccali et~al. 2014).  
The mean velocity and velocity dispersion of both the APOGEE giant sample
and the GIBS RCGs
are within 1~sigma of the mean velocity and velocity dispersion 
for the RRLs in the ($l$,$b$)=(3,$-$2.5) BRAVA-RR field (Kunder et~al. 2016, in prep).  Assuming the radial velocity 
distribution for the giants, red clump and RRLs is Gaussian, a star with $V_{GRV}$=$-$350~km~s$^{-1}$ 
is a four-sigma outlier in velocity space.  

The $\sim$9000~BRAVA giant sample \citep{kunder12}, $\sim$17000~ARGOS RCG 
sample \citep{freeman13} and the $\sim$1200~{\it Gaia}-ESO Survey bulge RCG 
sample \citep{rojasarriagada14} probe further from the Galactic plane than where
MACHO~176.18833.411 resides, so more contamination from e.g., the disk and halo would be expected,
but these large radial velocity surveys can also give an indication of how unusual it is for a star 
toward the bulge to have such a negative velocity.
Within the {\it Gaia}-ESO Survey bulge sample, there is no star with $V_{GRV}$ less 
than $-$350~km~s$^{-1}$ and $<$0.1\% of stars have $V_{GRV}$ less than $-$300~km~s$^{-1}$.
For both the BRAVA and ARGOS samples, $\sim$0.02\% of stars have $V_{GRV}$ less 
than $-$350~km~s$^{-1}$ and $\sim$0.1\% of stars have $V_{GRV}$ less than $-$300~km~s$^{-1}$.

\subsection{A halo star}
There are a few possibilities that can explain the presence of MACHO~176.18833.411.  
The most likely is that this star is a halo interloper that happens to be at the same distance
and location as where bulge RRLs reside.  This is supported by our orbital solution of 
MACHO~176.18833.411, which shows its $\rm R_{max}$ 
is mostly larger than 8~kpc, strongly suggesting 
this RRL is not confined to the bulge.  The time spent in the bulge is 145$\pm$100 Myr,
which corresponds to less than 15\% of the integration.
The $Z_{max}$ of $\sim$8~kpc suggests this star is not 
part of the disk, which is also consistent with its high velocity.  
This star is on an elliptical orbit (orbital ellipticity is 0.95 with a 
period of $\sim$140 Myr) and resembles a star in the halo.  
 
Numerical simulations which match the kinematic observations of the bulge well do 
not predict such high velocity bulge stars in this region of the sky \citep{shen10}, although 
we note that it is possible for significant radial velocity outliers to be present in any population.
Instead, high velocity RRLs have been found in the halo (e.g., $\sim$1.5\% of the local RRLs
in the Layden 1994 and Kollmeier et al. 2013 samples have 
$V_{GRV}$ less than $-$300~km~s$^{-1}$,
which is more than a factor of 10 larger than high velocity giants and RCGs found
in the bulge surveys discussed previously).  
Similarly within the halo globular cluster (GC) sample, 3\% have
$V_{GRV}$ less than $-$350~km~s$^{-1}$.

\begin{figure}[htb]
\includegraphics[width=1\hsize]{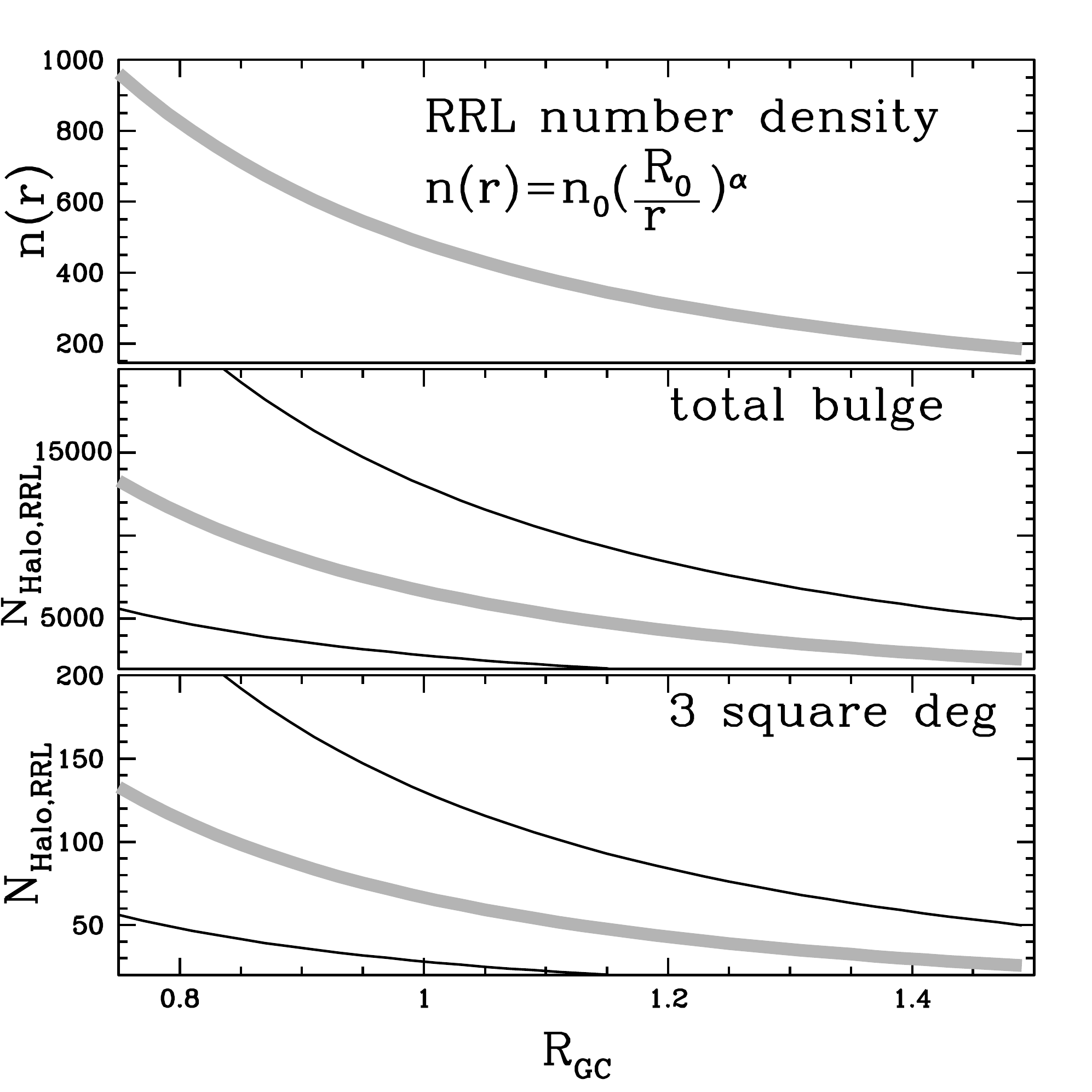}
\caption{{\sl Top:} The RRL number density as a function of galactocentric
distance, $R_{GC}$, in the 0.75 -- 1.5 kpc range using a power law derived from 
SDSS Stripe 82 RR Lyrae stars \citep{watkins09}, where ($\rm n_0$,$\rm R_0$, $\alpha$) = (0.26 kpc$^{-3}$, 23 kpc, 2.4). 
{\sl Middle:} The total number of halo RRLs, $\rm N_{halo,RRL}$, within the approximate
bulge volume as a function $R_{GC}$.  The thick grey lines indicates the
$N_{halo,RRL}$ assuming a bulge radius of 2~kpc and a minor to major axis ratio of 0.55;
the thin black lines indicate $\rm N_{halo,RRL}$ assuming a bulge radius of 1.5 and 2.5~kpc. 
{\sl Bottom:}  Same as the middle panel, for for a 3~square degree bulge field.
\label{numdensity}}
\end{figure}

If this RRL is a halo interloper, we presume additional halo 
contamination exists in the bulge RRL sample, with velocities indistinguishable
from bulge stars.  We would then expect at least a few percent of the RRLs located in
the direction toward the bulge to be halo constituents.  
To obtain a rough approximation of the presence of the halo, the relation
of the RRL density profile in the halo is extrapolated to the Galactic center \citep{watkins09}
as shown in Figure~\ref{numdensity}.  To find the total number 
of halo RRLs expected in our observations, we use this relation along with the 
following approximations:  (1) the bulge is a cylinder with
a radius of 2~kpc and a minor-to-major axial ratio of 0.55, and (2) the bulge
extends from $l = -$10 to $l=$+10 and $b = -$10 to $b=$+10 degrees, and so the AAOmega 3~square degree
field covers 3/314 of the Galactic bulge.  It follows, then, that 
$\sim$85$\pm$70 halo RRLs are expected with a Galactocentric distance of 0.9~kpc in a 
3~square degree field.  The uncertainty in the number of halo RRLs is a function
of the exact scale height and length of the bulge, and is illustrated in
Figure~\ref{numdensity}.  The total
number of OGLE RRLs in our 3~square degree field is $\sim$1400, and the Galactocentric
distance of these stars peaks at 0.8~kpc.  Therefore, assuming the OGLE sample 
of RRLs is relatively complete, $\sim$6\% of the OGLE RRLs stars toward ($l$,$b$)=(3,$-$2.5) are 
expected to be from the halo, and it is not unexpected for a few percent of
halo stars to exist so close, both in spatial and in distance distribution, to the Galactic center.
Proper motions of a larger sample of RRLs would be desirable to identify halo RRLs from
bulge RRLs. 

It is worth noting that the orbit of MACHO~176.18833.411 has 
its largest excursions perpendicular to the plane -- this extreme high velocity star is 
confined (statistically) vertically.  If the most 
extreme halo stars have an almost spheroidal zone of excursion, 
the (halo) population from where this star originated from is not the most extreme.
It may well be that this RRL is a relic from an earlier era, but likely not the earliest.

\subsection{A bulge star}
A second possibility is that MACHO~176.18833.411 is on the extreme edge of the 
bulge RRL radial velocity distribution.  This could be in line with the notion discussed 
by \citet{dekany13}, that the bulge RRLs follow a spheroidal, centrally 
concentrated distribution, as then eccentric orbits and therefore stars with large
radial velocities, would be expected.  
If the bulge RRLs follow a Gaussian radial velocity distribution with a $\sigma$=200~km~s$^{-1}$,
it would be expected to find a $V_{GRV}$$=$$-$350~km~s$^{-1}$ radial velocity star.  
Such a large velocity distribution is not observed 
for the 94 BRAVA-RR stars in this area of the sky, but although 
a Gaussian distribution is convenient, it is often wrong out on the tails, in a sense that the tails are too small.
It has also been shown that multiple RRLs populations exist in the bulge \citep{pietrukowicz14}, and so
there may be a small population of RRLs more kinematically hot than
the majority of bulge RRLs, and that MACHO~176.18833.411 is part of this more kinematically
hot component.  

Another motivation for MACHO~176.18833.411 being a member of the bulge RRL population
is that a similar proportion of high negative velocity stars are found in bulge stars 
exhibiting maser emissions \citep[in both OH/IR and SiO masers,][]{vanlangevelde92, fujii06}.  
However, it is unclear why these mass-losing infrared objects, typically Asymptotic Giant 
Branch (AGB) stars, which are thought to be much younger than 
RRLs, \citep[$\sim$1 to a few Gyr, ]{mouhcine02},  
would be more dynamically similar to bulge RRLs than red giants or RCGs.  

\subsection{A runaway star}
A third scenario to explain the velocity of MACHO~176.18833.411 is that it was
ejected from the Galactic center due to an interaction -- e.g., a star--binary 
or a star -- black hole collision, or ejected out of a globular cluster. 
It is unlikely that this star is a hypervelocity star (HVS), as its 
velocity is still well within the realm of being bound to the Milky Way \citep[e.g.,][]{kenyon08} 
MACHO~176.18833.411 may be a runaway star, though.  Evolved stars have been shown to 
be runaways \citep{kilic13}, as have horizontal branch stars \citep{pereira13}, although by far the
largest population of runaway stars currently known are younger, 
more massive stars \citep[e.g.][]{bromley09}.  Runaway stars are expected to dominate at low 
Galactic latitudes and are located preferentially in the direction of the Galactic center, between
$l$=325 -- 35 \citep[e.g.,][]{bromley09}.  The expected metallicity of runaways
is $\sim$$\rm [Fe/H]$$\sim$$-$1.5~dex, slightly more metal-rich than the halo
and comparable to MACHO~176.18833.411. 

\section{Conclusions}
In this paper, we take a detailed look at a high negative-velocity RRL observed
toward the Galactic bulge found in the BRAVA-RR survey.  Stars of such high 
velocity are rare among bulge giants, and the more precise distance of the RRL 
makes it possible to explore its origin in greater detail, by integrating its orbit. 
We argue that MACHO~176.18833.411 is
most likely a halo interloper, suggesting that contamination from halo stars 
is relevant when attempting to trace out the 
metal-poor tail of the bulge's metallicity distribution function \citep[e.g.,][]{howes14}.

\acknowledgements
We thank the Australian Astronomical Observatory, which have made these observations possible. 
This research was supported in part by the National Science Foundation under Grant No. NSF PHY11-25915.
This work was supported by Sonderforschungsbereich SFB 881 "The Milky 
Way System" (subprojects A4, A5, A8) of the German Research Foundation (DFG).
This work has been partially supported by the Polish Ministry of Science
and Higher Education through the program ``Ideas Plus" award No.
IdP2012 000162 to IS.  RMR acknowledges support from grant
AST-1413755  from the National Science Foundation.  C.I.J. gratefully acknowledges 
support from the Clay Fellowship, administered by the Smithsonian Astrophysical Observatory.

\clearpage

\end{document}